\documentclass[a4paper,11pt]{article}
\usepackage{jinstpub} 
\usepackage{lineno}
\usepackage[percent]{overpic}
\usepackage{acro}
\usepackage{xspace}
\usepackage{hyperref}
\usepackage{academicons} 
\usepackage{svg} 
\usepackage{siunitx}
\usepackage{rotating} 
\usepackage{subcaption}

\DeclareAcronym{lgad}{
  short = LGAD,
  long  = Low Gain Avalanche Diode,
  tag = abbrev
}

\DeclareAcronym{lhc}{
  short = LHC,
  long  = Large Hadron Collider,
  tag = abbrev
}

\DeclareAcronym{pcb}{
  short = PCB,
  long  = printed circuit board,
  tag = abbrev
}

\DeclareAcronym{QA}{
  short = QA,
  long  = quality assurance,
  tag = abbrev
}

\DeclareAcronym{QC}{
  short = QC,
  long  = quality control,
  tag = abbrev
}

\DeclareAcronym{IV}{
  short = I--V,
  long  = current--voltage,
  tag = abbrev
}

\DeclareAcronym{CV}{
  short = C--V,
  long  = capacitance--voltage,
  tag = abbrev
}

\DeclareAcronym{ffc}{
  short = FFC,
  long  = flexible flat cable,
  tag = abbrev
}

\DeclareAcronym{smu}{
  short = SMU,
  long  = source-measure unit,
  tag = abbrev
}

\newcommand{\um}{\si{\micro\meter}}

\newcommand{\mmsq}{\si{\milli\meter\squared}}

\usepackage{bm,epstopdf,natbib,hyperref,color,verbatim,multirow,bm,tikz,xcolor}
\hypersetup{colorlinks=true,urlcolor=blue,citecolor=blue,linkcolor=blue,menucolor=blue,anchorcolor=blue,filecolor=blue}
\everymath{\displaystyle}
\definecolor{lime}{HTML}{A6CE39}
\DeclareRobustCommand{\orcidicon}{
  \begin{tikzpicture}
    \draw[lime, fill=lime] (0,0) 
    circle [radius=0.16] 
    node[white] {{\fontfamily{qag}\selectfont \tiny ID}};
    \draw[white, fill=white] (-0.0625,0.095) 
    circle [radius=0.007];
  \end{tikzpicture}
  \hspace{-3mm}
}
\foreach \x in {A, ..., Z}{\expandafter\xdef\csname orcid\x\endcsname{\noexpand\href{https://orcid.org/\csname orcidauthor\x\endcsname}
  {\noexpand\orcidicon}}
}

\title{\boldmath Development of a system for testing full-size CMS LGAD sensors}

\author[a]{Kyungmin Lee\hspace{-1mm}\orcidC{},}
\author[a]{Hoyong Jeong\hspace{-1mm}\orcidB{},}
\author[b]{Junho Kim\hspace{-1mm}\orcidD{},}
\author[b]{Seokhyeon Lee\hspace{-1mm}\orcidF{},}
\author[a]{Jaebak Kim*\hspace{-1mm}\orcidE{},}
\author[a]{Jae Hyeok Yoo*\hspace{-1mm}\orcidA{}}

\affiliation[a]{Department of Physics, Korea University, Seoul, 02841, Korea}
\affiliation[b]{Department of Physics, Kyungpook National University, Daegu, 41566, Korea}

\emailAdd{railroad@korea.ac.kr}
\emailAdd{jaebak@korea.ac.kr}
\emailAdd{jaehyeokyoo@korea.ac.kr}

\abstract{ 
Low-Gain Avalanche Diode (LGAD) sensors, offering timing resolutions of
the order of tens of picoseconds, are being widely adopted in particle
physics experiments and related applications.
As these applications scale to large numbers of sensors with varying
pixel geometries, conventional manual characterization techniques become
inadequate for large-scale quality control.
We present a modular probe card system for automated electrical
characterization of pixelated LGAD sensors, consisting of a probe card,
a switching board, precision measurement instruments, and control
software.
The system supports flexible pixel selection and measurement.  
Its performance is demonstrated through current-voltage (I-V) and capacitance-voltage (C-V)
measurements of a $16 \times 16$ LGAD array.
A rapid row-wise I-V scan of the full array is completed in
approximately 20 minutes, while a pixel-by-pixel I-V scan from
0 to 300~V with a 1~V step requires about 340 minutes.
The switching matrix introduces less than 1~nA of leakage current even
in a conservative worst-case configuration, remaining small compared
with the leakage current of a normal LGAD pixel.
The modular architecture and automation capability make the system a
practical and scalable solution for large-scale LGAD sensor quality
control and distributed testing environments.
}

\begin{document}

\maketitle

\flushbottom

\section{Introduction}\label{sec:intro}

The Low-Gain Avalanche Diode (LGAD) is a silicon detector capable of providing 
precise timing resolution 
and resistance to high radiation fluence
~\cite{LGADIntro}.
LGADs have been developed to meet the stringent timing requirements of 
high-luminosity collider environments. 
In these settings, extreme pileup conditions make spatial tracking alone 
insufficient for accurate track--vertex association. 
By integrating high-precision temporal information with spatial 
coordinates---a technique known as four-dimensional (4D) tracking---LGADs 
substantially improve pattern recognition and provide powerful capabilities 
for pileup mitigation.

The combination of moderate internal gain and inherent radiation 
hardness makes LGADs the preferred sensor technology for precision 
timing applications in current and future particle physics experiments.
They have been used in the CMS-TOTEM Precision Proton Spectrometer since 2017, 
with sensors of 50~\um{} thickness and pixel sizes of 
$0.5 \times 6$~\mmsq{} and $1 \times 3$~\mmsq{}, 
achieving a timing resolution of 35~ps~\cite{CMSTOTEM, CMSTOTEM2, CMSTOTEM3}.
LGADs are planned for deployment in the High-Luminosity Large Hadron Collider
era at the ATLAS and CMS experiments, where the sensors have a thickness of 
50~\um{} with $1.3 \times 1.3$~\mmsq{} pixels and approximately 50~ps timing 
resolution~\cite{ATLASLGAD, CMSLGAD, CMSLGAD2}. 
They are being investigated for use in the HADES fixed-target experiment~\cite{HADES},
where a 50~\um{} thick strip sensor with a timing resolution of 47~ps was studied.
There are also variants of LGADs under development, such as the 
AC-LGAD~\cite{ACLGAD}, which enables finer segmentation or a larger fill factor. 
LGADs have also been selected for the time-of-flight system of the ePIC 
experiment, featuring sensors with thicknesses below 30~\um{}, pixel sizes of 
$1.6 \times 1.6$~\mmsq{}, and an approximate timing resolution of 
20~ps~\cite{ePIC}.
Applications of LGADs in space experiments and medical fields are also being 
investigated, leveraging their high timing precision~\cite{SPACE, MEDICAL}.   

For LGAD sensors to be deployed in experiments and applications, their 
performance must be carefully characterized, and each device must undergo 
strict \ac{QC} procedures.                                                    
This is particularly critical for next-generation experiments that will rely  
on thousands of LGAD sensors, such as those planned for the ATLAS and CMS
experiments.
Conventional characterization techniques are often time-consuming and 
resource-intensive, making them insufficient for these next-generation 
experiments.

Previous studies have reported \ac{QC} of LGAD sensors using probe cards
and switching boards developed for the ATLAS experiment~\cite{ATLASSWITCH, ATLASSWITCH2}.
These systems enable automated \ac{IV} and \ac{CV} measurements of
$5 \times 5$ and $15 \times 15$ LGAD arrays with noise levels below 10~pA.
Their switching architectures are primarily optimized for sequential
single-pixel measurements, which are well-suited for pixel-by-pixel
quality control.
For larger pixel arrays and a broader range of test scenarios, additional channel-routing flexibility can be beneficial. Such flexibility is particularly useful for multi-pixel measurements, row- or column-wise scans, and inter-pixel studies, where the inter-pixel configuration enables measurements of the electrical characteristics between selected pixels.
A modular hardware structure can also simplify scaling, maintenance,
and reconfiguration of the system.
Motivated by these requirements, we developed a modular probe card
system incorporating a switching architecture that enables automated
characterization of large pixelated LGAD sensors with arbitrary pixel
selection.

The remainder of the paper is organized as follows.
Section~\ref{sec:system} introduces the LGAD sensor from the perspective
of electrical characterization and provides an overview of the evaluation
system and its measurement requirements.
The individual subsystems are then described in detail: the probe card in
Section~\ref{sec:probecard}, the mechanical structure and alignment system
in Section~\ref{sec:mechanics}, the switching board in
Section~\ref{sec:swboard}, and the control software in
Section~\ref{sec:software}.
The overall system performance is presented in
Section~\ref{sec:performance}, followed by the conclusions in
Section~\ref{sec:summary}.

\section{System design for LGAD sensor evaluation}\label{sec:system}
To meet the demands of large-scale LGAD sensor quality control, 
we developed a modular probe card system for reliable and efficient 
electrical characterization of pixelated LGAD sensors with large 
pixel arrays, enabling pixel-by-pixel and multi-pixel \ac{IV} and \ac{CV} 
measurements.
This section first describes the LGAD sensor structure from the perspective 
of electrical characterization. The measurement objectives and system 
requirements are then introduced, followed by an overview of the evaluation 
system architecture.

\begin{figure}[h]
\begin{center}
\includegraphics[width=0.99\linewidth]{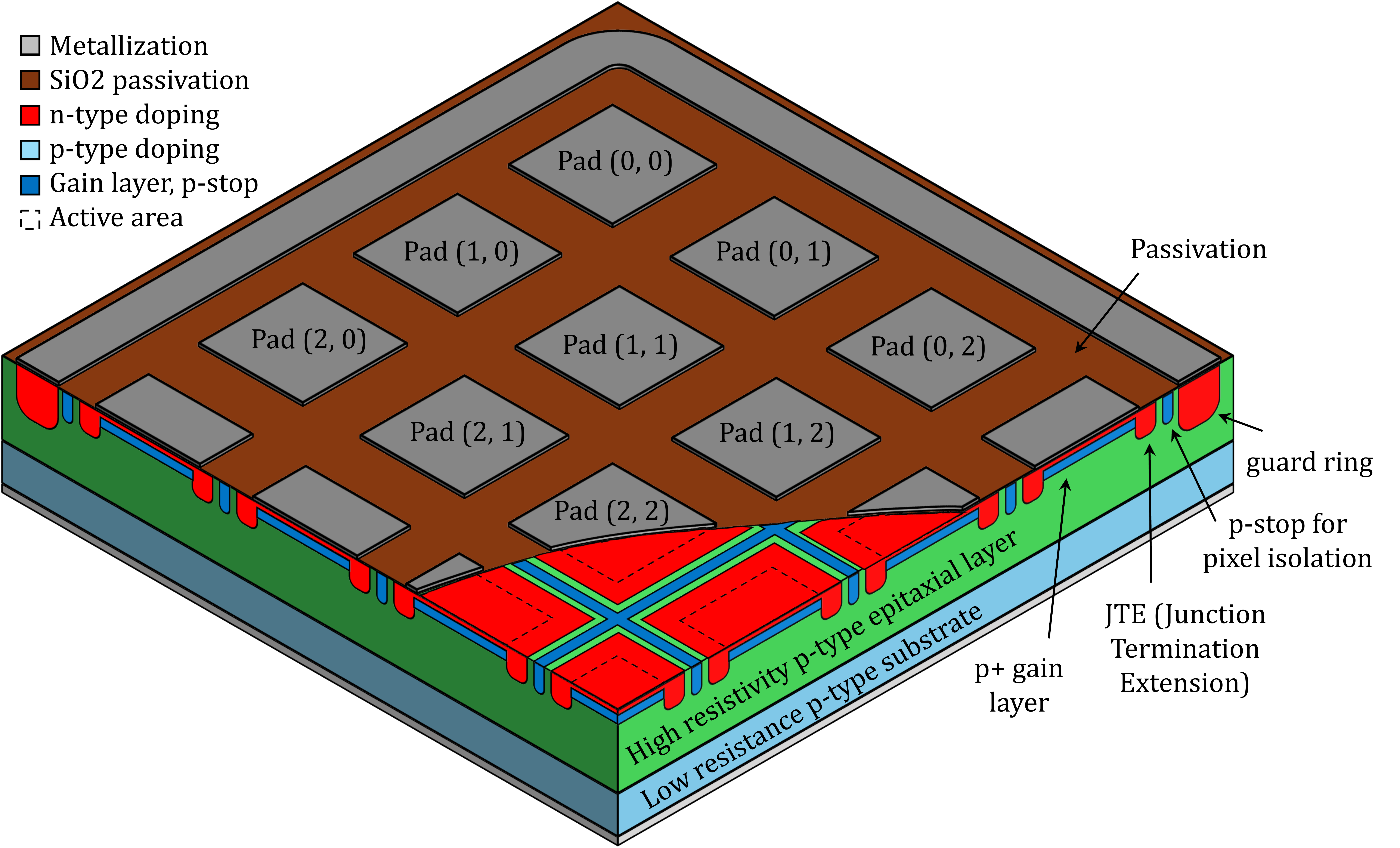}
\end{center}
\caption{Schematic diagram of a typical LGAD sensor structure, 
showing a partial $4 \times 4$ array with a cross-sectional view. 
The cross-section reveals the internal layer structure from bottom 
to top: the low-resistivity p-type substrate serving as the ohmic 
contact, the high-resistivity p-type epitaxial layer forming the 
sensitive volume, and the p$^{+}$ gain layer and n$^{++}$ readout 
electrode beneath each metal contact pad. 
JTE structures are implemented at the periphery of each pixel to 
prevent premature breakdown, and adjacent pixels and the guard ring 
are electrically isolated by p-stop structures. 
Guard rings are implemented at the sensor periphery to 
provide additional edge termination.}
\label{fig:LGADillustration}
\end{figure}

Figure~\ref{fig:LGADillustration} shows a schematic diagram of a typical 
LGAD structure. The device is based on a modified p--i--n diode geometry 
that incorporates a p$^{+}$ gain layer beneath the n$^{++}$ readout electrode. 
Under reverse bias, the gain layer creates a localized region of high 
electric field. Electrons drifting into this region undergo controlled 
impact ionization, producing moderate internal charge multiplication with 
a typical gain of 10--30. This internal gain compensates for the reduced 
signal charge in thin sensors and enables precise timing measurements.

The sensitive volume of the sensor is formed by a high-resistivity 
epitaxial layer with a thickness of approximately 50~\um{}.
A heavily doped p$^{++}$ substrate on the backside serves as the 
ohmic contact for reverse bias.
On the top surface, each pixel incorporates a p$^{+}$ gain layer 
and an n$^{++}$ readout electrode, with a metal contact pad formed 
above it. Junction Termination Extension (JTE) structures are implemented at the periphery of each 
pixel to prevent premature breakdown.
Adjacent pixels and the guard ring are electrically isolated by 
p-stop structures.
At the sensor periphery, guard rings provide additional 
edge termination.

In this work, we consider LGAD sensors developed for the CMS Endcap Timing 
Layer, which consist of a $16 \times 16$ pixel array \cite{CMS:2667167}. 
The key electrical characteristics of each pixel are governed 
by the device structure and must be measured individually to 
verify fabrication uniformity and ensure stable operation 
across the full array.

The primary objective of LGAD sensor characterization is to determine the 
safe operating voltage range of each pixel while ensuring sufficient internal
gain for precise timing measurements. This is achieved through \ac{IV} 
measurements, which provide information on breakdown voltage and leakage 
current. In addition, \ac{CV} measurements are used to study the depletion 
behavior of the sensor and to assess the uniformity of the bulk and gain 
layer properties. These measurements define the operating conditions of 
the sensor and directly influence the achievable internal gain and charge
collection efficiency, both of which determine the timing performance of 
LGAD sensors.

Based on these measurement objectives, several system requirements are identified for large-scale LGAD sensor evaluation:
\begin{itemize}
\item Precise and reliable pixel-level electrical characterization with low noise contribution from the readout chain
\item Flexible pixel selection capability supporting single-pixel, multi-pixel, and inter-pixel measurement configurations
\item Rapid measurement throughput suitable for large-scale quality control during mass production
\item Cost-effective and scalable architecture for deployment across multiple institutes and testing environments
\end{itemize}

Accurate characterization requires pixel-by-pixel measurements in which the 
electrical response of each pixel is measured individually. During each 
measurement, the pixel under test must be connected to the measurement 
instrument while all other pixels are held at a well-defined potential, 
typically ground. This configuration suppresses parasitic contributions 
from neighboring pixels, such as electrical cross-talk and leakage current 
sharing, ensuring that the measured signal originates only from the 
selected pixel.

For sensors with a small number of pixels, such measurements can be performed 
using conventional probe stations equipped with needle probes. However, 
for sensors containing hundreds of pixels, such as the $16 \times 16$ array 
considered in this work, manual probing becomes impractical and difficult
to scale. A dedicated probing and switching system is therefore required.

The probe card establishes electrical contact to all pixels through a 
pogo-pin array matched to the sensor geometry. A switching board then 
selectively routes the signal from a chosen pixel to the measurement 
instruments while grounding all other pixels. This approach enables 
pixel-by-pixel measurements without changing the physical probing contact and 
allows systematic characterization using a limited number of instrument 
channels.

To implement this concept, we developed an LGAD sensor evaluation system 
consisting of four main components: a probe card, a switching board, 
measurement instruments, and control software. The measurement instruments 
include a \ac{smu} and a picoammeter for \ac{IV} measurements, as well as 
an inductance--capacitance--resistance (LCR) meter for \ac{CV} measurements. 
The control software coordinates the switching configuration and the 
operation of the instruments, enabling automated scan sequences and 
systematic data acquisition.

\begin{figure}[h]
\begin{center}
\includegraphics[width=0.9\linewidth]{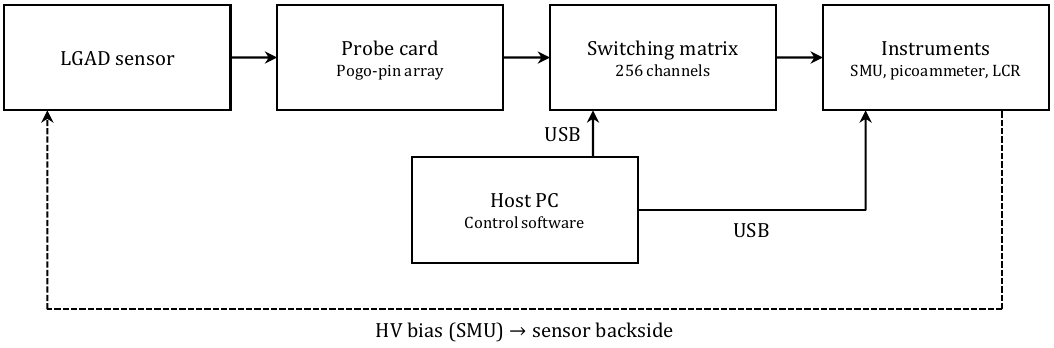}
\end{center}
\caption{Block diagram of the probe card system architecture, 
consisting of the LGAD sensor, probe card, switching matrix, 
and measurement instruments. The host PC coordinates all 
components via USB interfaces. The SMU applies 
reverse bias to the sensor backside (dashed line) while pixel 
currents are routed through the switching matrix.}
\label{fig:dummy}
\end{figure}

Together, these components form a modular system that enables automated 
pixel-by-pixel electrical characterization of large pixelated LGAD sensors.
An overview of the system architecture and the interaction between its 
components is illustrated in Fig.~\ref{fig:dummy}. The following sections 
describe each subsystem in detail.

\section{Probe card}\label{sec:probecard}
A probe card was developed to establish simultaneous electrical contact to all pixels of a $16 \times 16$ LGAD array for automated electrical characterizations.
The pogo-pin layout was designed to match the geometry of LGAD sensors developed for 
minimum-ionizing particle (MIP) timing detectors of the CMS Endcap Timing Layer (ETL).
The spacing between adjacent pogo pins is 1.3~mm, corresponding to the pixel pitch of the target sensor.
In addition to the $16 \times 16$ signal pins, 15 extra pogo pins were 
placed along one side of the array to contact the guard ring pads. 

Pogo pins were selected as the contacting elements because they provide
mechanically robust and repeatable contact over many measurement cycles.
Compared with cantilever-type probes, pogo pins are less susceptible to
permanent deformation and offer more stable contact conditions during
repeated \ac{IV} and \ac{CV} scans, thereby reducing the risk of contact
failure or sensor damage.
The pogo pins have a spring stroke of 0.5~mm, and the full span of the
pin array is 19.5~mm. These parameters determine the allowable tilt 
tolerance between the probe card and the sensor surface, as discussed 
in Section~\ref{sec:mechanics}.

Electrical connections from the probe card to the switching board are
provided through three 96-pin \ac{ffc} connectors, which carry the 256
pixel signal lines; unused connector pins are left unconnected.
An \ac{ffc} adapter board reorganizes these lines into sixteen 16-channel
groups for connection to the motherboard of the switching system.
The 15 guard ring lines are routed separately through a 20-pin
\ac{ffc} connector and are directly connected to ground.
This separation simplifies the routing of the signal channels while
keeping the guard ring at a fixed reference potential during
measurements.
Through this connector scheme, each pixel signal can be selectively
routed by the switching board either to a measurement bus or to ground,
thereby enabling flexible \ac{IV} and \ac{CV} configurations without
changing the physical probing contact.

\begin{figure}[h]
\begin{center}
\includegraphics[width=0.7\linewidth]{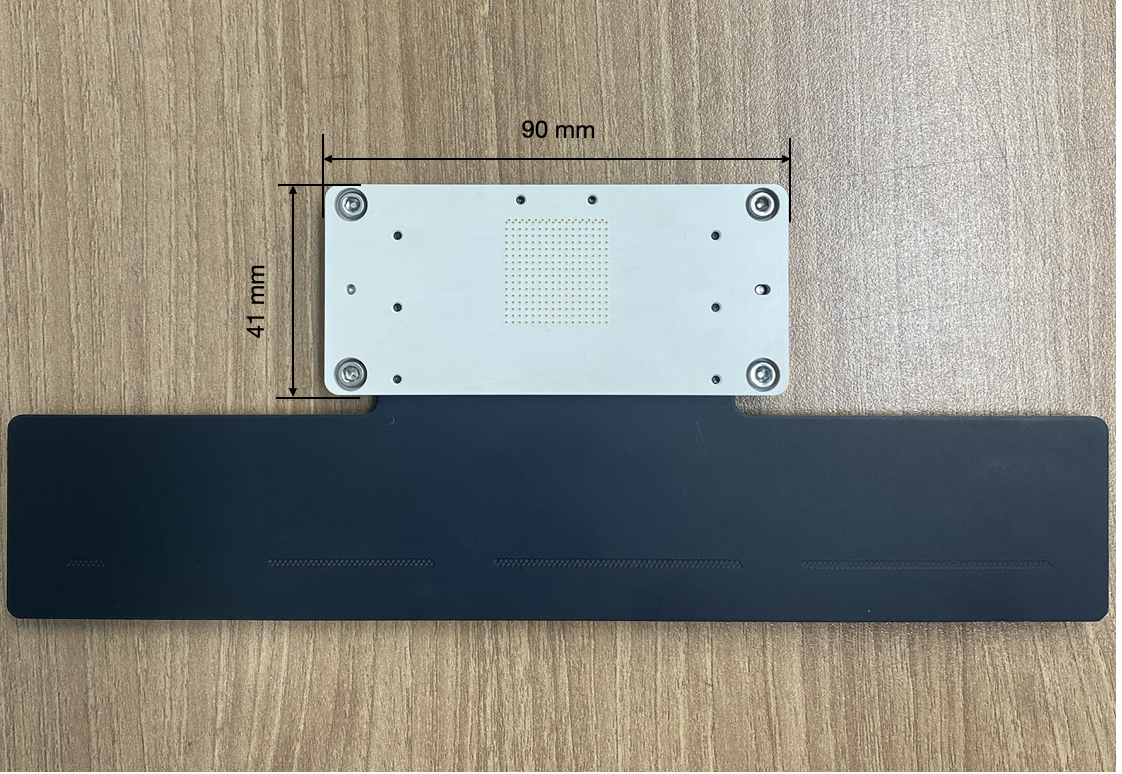}
\end{center}
\caption{Fabricated probe card.
The polymer head (90~mm $\times$ 41~mm) houses the central
$16 \times 16$ pogo-pin array for pixel contact and 15 additional pogo
pins for the guard ring pads.
The elongated \ac{pcb} section supports the \ac{ffc} connectors used to
route the pixel signals from the probe card to the switching board on the reverse side.}
\label{fig:probe_card}
\end{figure}

The probe card shown in Fig.~\ref{fig:probe_card} was designed and fabricated 
in collaboration with EQENG, a probe card manufacturer based in Korea. 
The fabricated probe card was successfully used for automated \ac{IV} and \ac{CV} measurements of $16 \times 16$ LGAD sensors, as presented in Section~\ref{sec:performance}.

\section{Mechanics and alignment}\label{sec:mechanics}

To establish simultaneous and reliable contact across the full
$16 \times 16$ pogo-pin array, precise control of the lateral position,
tilt, and vertical compression between the probe card and the sensor is
required.
A dedicated mechanical system was therefore developed for the sensor
alignment and probing.
The supporting structure is built from a modular aluminum profile frame,
while fine positioning and angular adjustments are provided by precision
motion stages.
The sensor is mounted on a vacuum chuck to ensure stable positioning
during the probing procedure.

\begin{figure}[h]
\begin{center}
\includegraphics[width=0.8\linewidth]{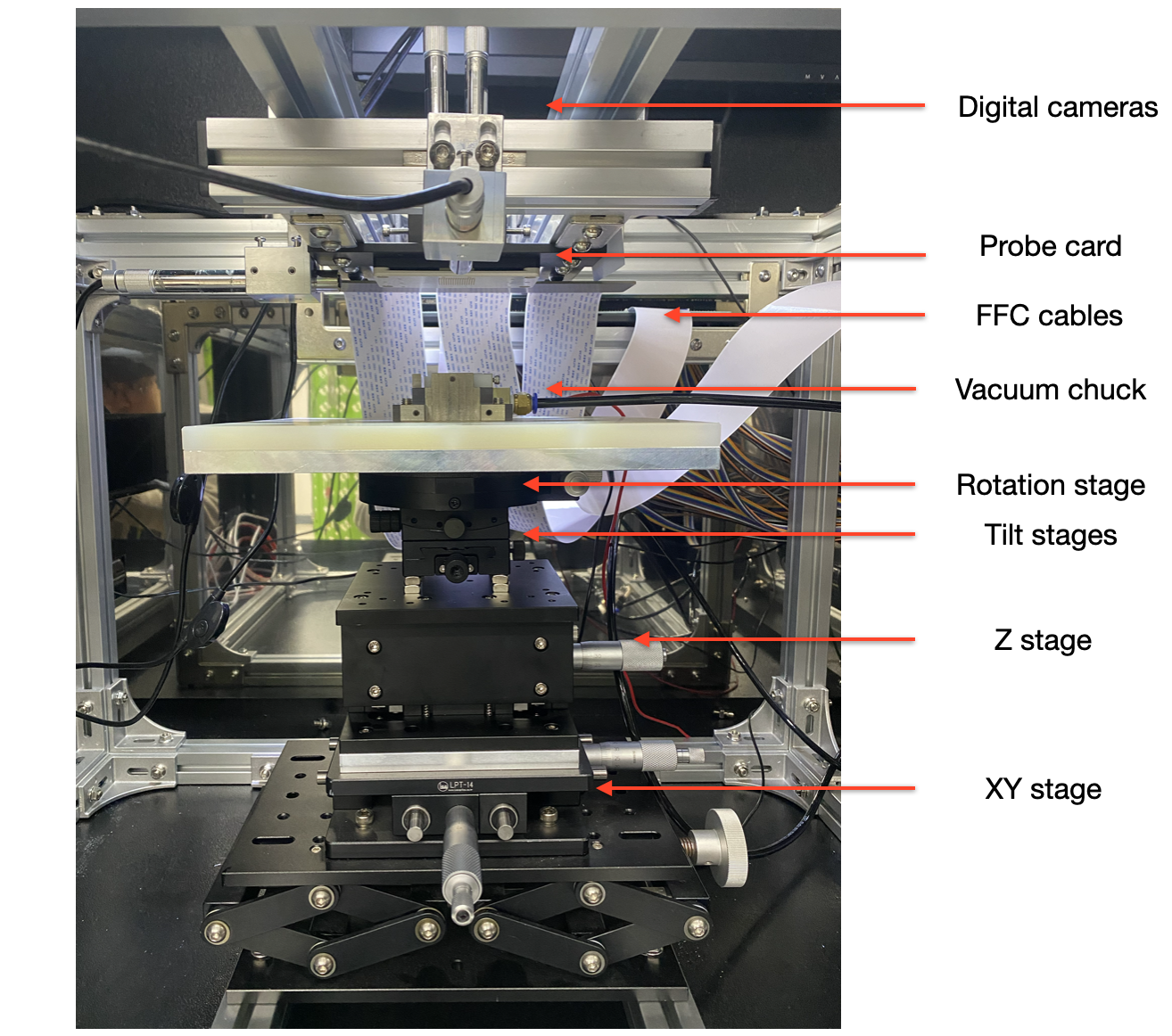}
\end{center}
\caption{Probing setup used for precise alignment
between the probe card and the sensor. The upper assembly supports the
probe card and alignment cameras, while the lower assembly carries the
sensor on a vacuum chuck together with the multi-axis positioning
stages. \ac{ffc} cables connect the probe card to the switching board.}
\label{fig:jig_system}
\end{figure}

The jig system, shown in Fig.~\ref{fig:jig_system}, consists of an upper
assembly that supports the probe card and a lower assembly that carries the sensor on the vacuum chuck together with the positioning stages.
This separation allows the probe card to remain mechanically stable
while the sensor position is adjusted with respect to it.
Dedicated holders are mounted on the frame for auxiliary components,
including the alignment cameras and the \ac{ffc} adapters.
These components are fixed directly to the aluminum profiles to maintain
their relative position with respect to the probe card and sensor during
alignment and measurement.

For alignment, two reference holes in the probe card are matched to
corresponding patterns on the sensor.
Two top-view cameras are used to monitor these alignment marks and to
adjust the lateral position and in-plane rotation between the probe card
and the sensor.
To monitor the relative tilt, two additional side-view cameras are
placed along the x and y directions of the sensor plane.
Alignment is performed by first correcting the x--y position and
rotation using the top-view images, then adjusting the tilt using the
side-view images, and finally approaching the sensor in the z direction
to achieve controlled compression of the pogo pins.

The stage specifications were selected based on the alignment tolerances
imposed by the sensor pad geometry and the pogo-pin contact conditions.
XY, Z, tilt, and rotation stages with a positioning resolution of
10~\um{} are employed. This resolution is well below the characteristic dimensions
of the sensor contact pads, which have typical dimensions of approximately
$200~\um{} \times 100~\um{}$. This resolution also satisfies the vertical direction requirement that is set by
the pogo-pin spring stroke of about 500~\um{}, which determines the
allowable range for controlled compression during probing.

Based on the pogo-pin stroke of 0.5~mm and the full pin-array span of
19.5~mm (Section~\ref{sec:probecard}), the height variation across the
array must remain within the available spring stroke to ensure that all
pins can make contact without over-compression.
This condition corresponds to an allowable tilt of approximately
\ang{1.4} between the probe card and the sensor surface.
To satisfy this requirement, a tilt stage with an angular resolution of
\ang{0.3} is used in the probing setup.

A similar geometric consideration applies to the in-plane rotation.
From the pad dimensions and the full lateral extent of the sensor, an
allowable rotational tolerance of approximately \ang{0.5} is sufficient
to keep the contact displacement within the pad area.

The complete probing setup is operated inside a light-tight enclosure
during measurements, due to visible-wavelength photons being absorbed in the silicon sensor that generate
electron--hole pairs, producing photocurrent that can distort both
\ac{IV} and \ac{CV} measurements.
The enclosure therefore suppresses photo-induced current and ensures
that the measured response reflects the intrinsic electrical properties
of the sensor.

\section{Switching board}\label{sec:swboard}

To enable automated \ac{IV} and \ac{CV} measurements of LGAD sensors with
a $16 \times 16$ pixel array, a dedicated switching board was developed.
Its primary function is to connect selected pixels to the external
measurement instruments while holding all non-selected pixels at ground
potential.
This configuration suppresses unwanted contributions from neighboring
pixels and enables systematic pixel-by-pixel characterization without
changing the physical probing contact or manually reconnecting cables.
In addition, the switching architecture supports flexible measurement
configurations beyond single-pixel access, as described below.

The switching system is implemented as a modular matrix providing
256 independent channels.
It consists of 16 unit boards, each handling 16 channels, mounted on a single motherboard.
Each channel corresponds to one LGAD pixel, allowing the full
$16 \times 16$ sensor array to be addressed through a common hardware
platform.
This modular organization simplifies assembly, maintenance, and possible
future adaptation to other sensor layouts.

The overall switching concept was inspired by previously reported
multi-channel LGAD switching schemes~\cite{ATLASSWITCH}, in which switches
and multiplexers are combined to access individual pixels.
In the present design, a simplified architecture based solely on analog
switches was adopted and combined with four independent measurement
buses.
This approach increases routing flexibility for multi-pixel and inter-pixel
measurements, at the cost of introducing a small additional offset or
leakage contribution from the shared signal paths.
As discussed in Section~\ref{sec:performance}, this contribution was
found to be small compared with the leakage current typically observed for functional LGAD pixels.

Analog switches (TMUX1134\footnote{\url{https://www.ti.com/product/ko-kr/TMUX1134}}, Texas Instruments) were selected as the core
switching elements.
Each device integrates four bidirectional single-pole double-throw
(SPDT) switches in a single package and provides low on-resistance with
a typical channel leakage current of about 50~pA.
The SPDT configuration is particularly useful because it allows each
pixel either to be routed to a measurement bus or to be tied to ground.
This ensures that only the selected pixel contributes to the measured
signal, while all other pixels remain at a well-defined reference
potential during the measurement.

\begin{figure}[htbp]
    \centering
    \begin{subfigure}[b]{0.56\textwidth}
        \centering
        \includegraphics[width=\linewidth]{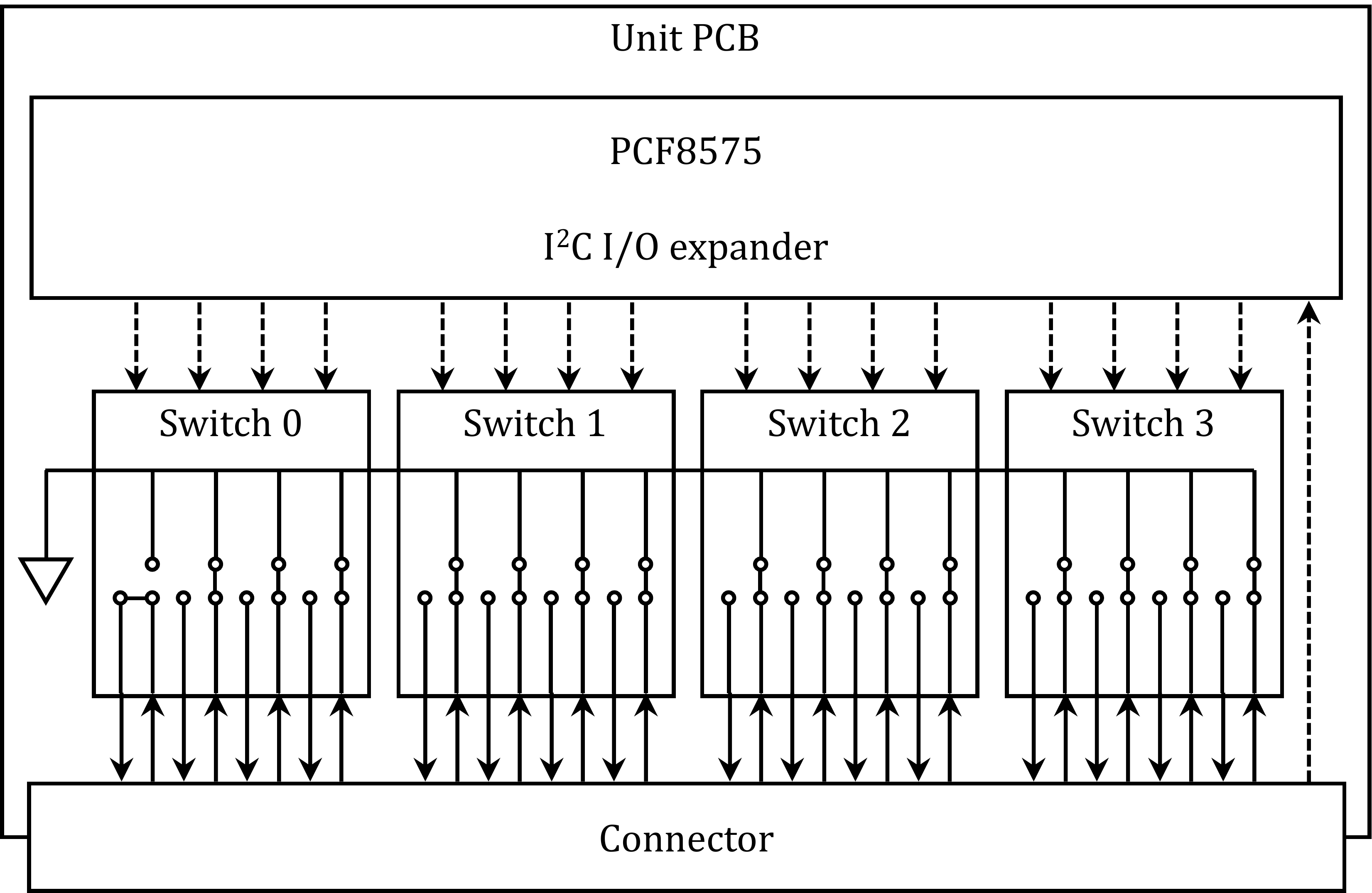}
        \caption{}
        \label{fig:unit_schematic}
    \end{subfigure}
    \hfill
    \begin{subfigure}[b]{0.42\textwidth}
        \centering
        \includegraphics[width=\textwidth]{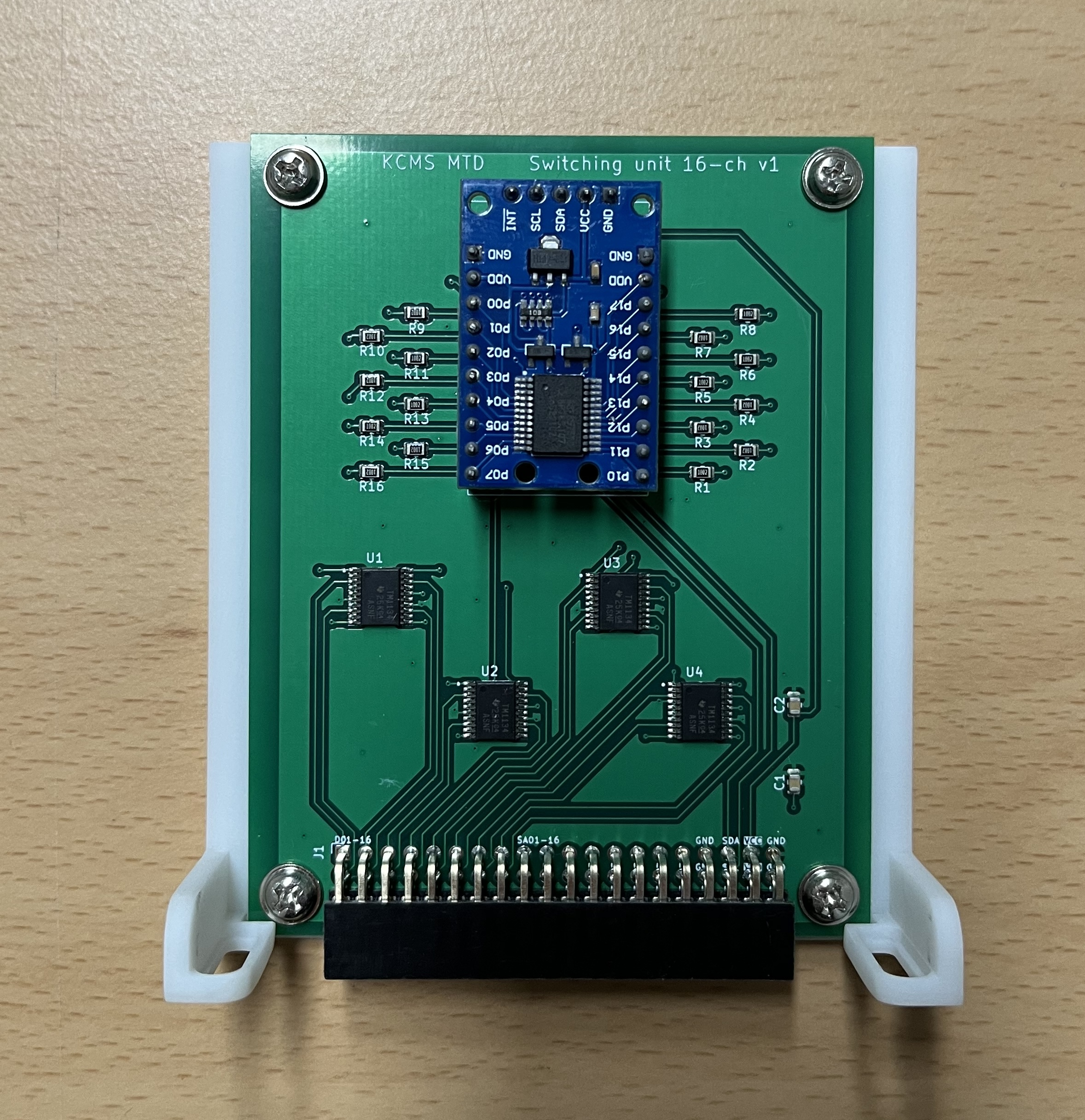}
        \caption{}
        \label{fig:unit_photo}
    \end{subfigure}
    \caption{Unit board (a) Schematic diagram. One unit board integrates four
    4-channel analog switches together with a PCF8575 I/O expander for
    digital control via I2C. (b) Fabricated unit board.}
    \label{fig:unit}
\end{figure}

A unit board serves as the basic building block of the switching matrix.
Each unit board contains four TMUX1134 chips, corresponding to
16 independent switching channels.
For each channel, the probe-card signal line is connected to the common
terminal of the switch, while the two switched outputs are connected to
a measurement bus and to ground, respectively.
This unit-level modularity simplifies the \ac{pcb} layout, keeps the
sensitive analog paths short, and allows individual boards to be
replaced or reconfigured when needed.
The schematic diagram and a photograph of the fabricated unit board 
are shown in Fig.~\ref{fig:unit}.

\begin{figure}[htbp]
    \centering
    \begin{subfigure}[b]{0.56\textwidth}
        \centering
        \includegraphics[width=\textwidth]{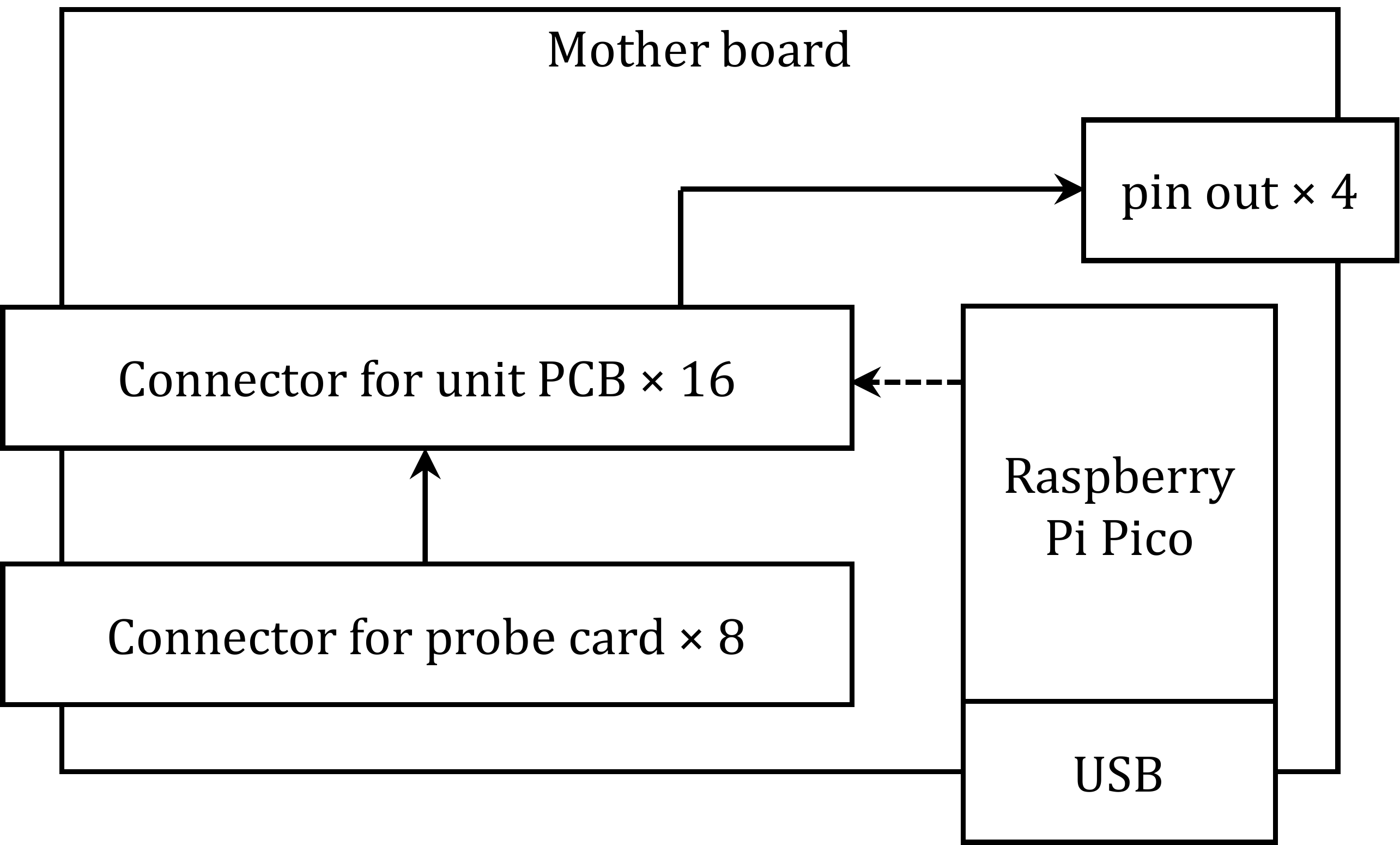}
        \caption{}
        \label{fig:mb_sch}
    \end{subfigure}
    \hfill
    \begin{subfigure}[b]{0.42\textwidth}
        \centering
        \includegraphics[width=\textwidth]{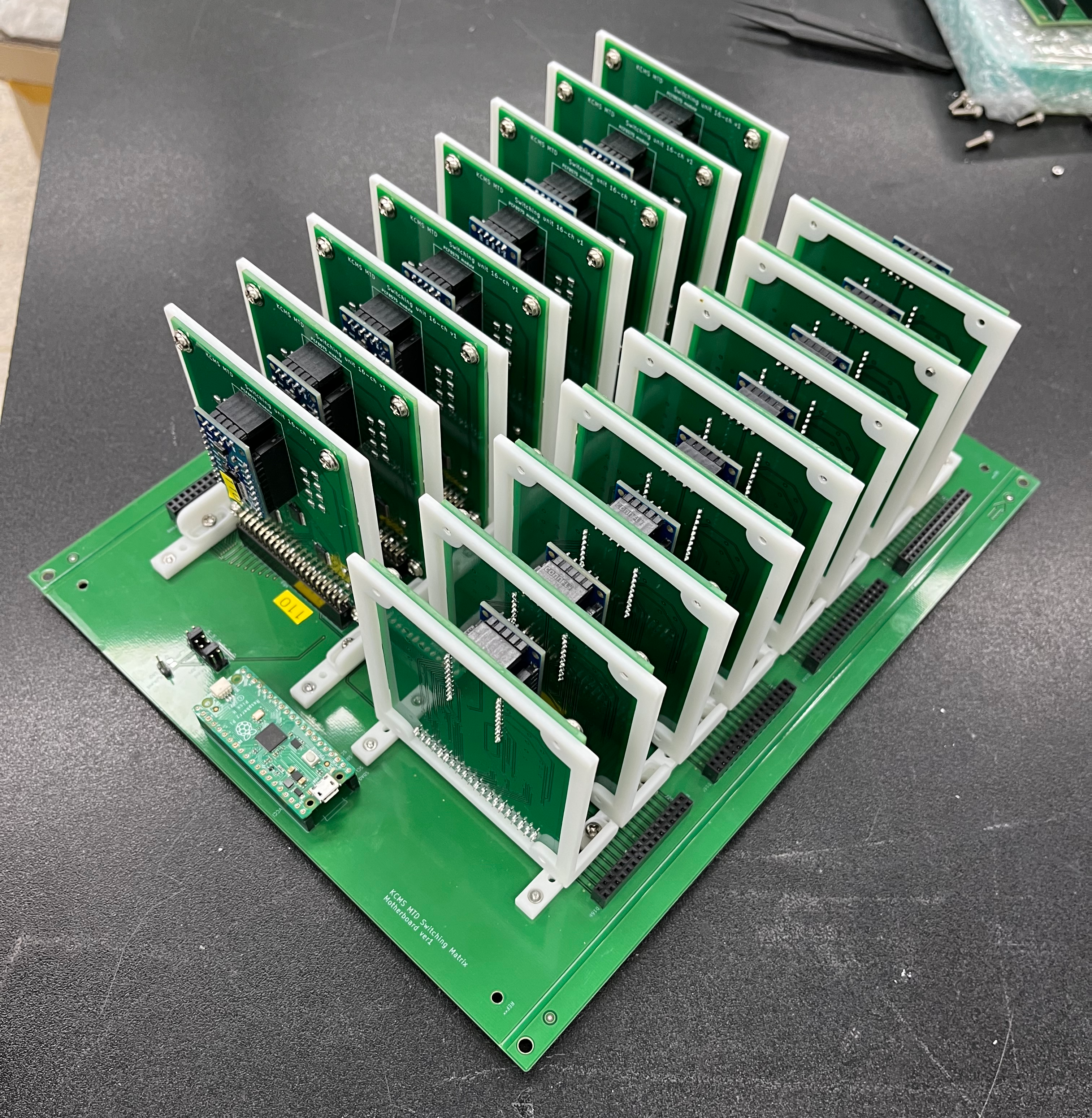}
        \caption{}
        \label{fig:mb_photo}
    \end{subfigure}
    \caption{Motherboard of the switching system.
    (a) Schematic diagram of the motherboard, illustrating the connections
    between the probe card, the unit \ac{pcb}s, the four output pins to the
    external measurement instruments, and the Raspberry Pi Pico
    microcontroller used for switch control and USB communication with the
    host PC.
    (b) Assembled switching system, showing the modular
    implementation with the unit \ac{pcb}s mounted vertically on the
    motherboard.}
    \label{fig:mb}
\end{figure}

The motherboard, shown in Fig.~\ref{fig:mb}, hosts 16 unit boards and
provides a total of 256 switching channels.
It distributes the signal lines from the probe card together with the
common ground, digital control signals, and power required by the unit
boards.
The 256 pixel signals are transferred from the probe card through three
96-pin \ac{ffc} connectors to an \ac{ffc} adapter board, which
reorganizes them into sixteen 16-channel groups.
These groups are then connected to the motherboard through eight
32-pin connectors, each serving two 16-channel groups.
The motherboard routes the selected signals to four independent
measurement buses connected to the external instruments.

This four-bus architecture supports several measurement modes,
including single-pixel, multi-pixel, row-wise, and column-wise
configurations.
Because each channel can be switched independently and adjacent pixels
are assigned to different buses, arbitrary pixel selection and
simultaneous multi-pixel access are both possible while the remaining
pixels are held at ground.
This flexibility is useful for routine pixel-by-pixel quality control,
rapid pre-scans, and inter-pixel measurements, in which neighboring
pixels can be connected to different buses so that the instrument probes
the electrical path between them.

To control the large number of switching channels, each unit board
integrates a PCF8575\footnote{\url{https://www.ti.com/product/PCF8575}} I/O expander together with the analog switches.
Each PCF8575 provides 16 digital output lines for controlling the
corresponding 16 switching channels via the I2C protocol.
A Raspberry Pi Pico based on the RP2040\footnote{\url{https://www.raspberrypi.com/products/rp2040/}} microcontroller serves as the
main controller, communicating with the PCF8575 devices via I2C and
with the host PC through a USB interface.
Because the RP2040 provides two independent I2C controllers and the
PCF8575 supports up to eight selectable addresses per bus, all 16 unit
boards can be controlled with a single microcontroller.

Control software running on the host PC coordinates the switching
configuration and synchronizes it with the external measurement
instruments.
For \ac{IV} measurements, a \ac{smu}\footnote{\url{https://www.tek.com/ko/products/keithley/source-measure-units/2400-standard-series-sourcemeter}} 
and a picoammeter\footnote{\url{https://www.tek.com/ko/products/keithley/low-level-sensitive-and-specialty-instruments/series-6400-picoammeters}} 
are used.
The \ac{smu} applies reverse bias to the backside of the sensor and
monitors the total current of the full array, while the picoammeter
measures the current from the selected pixel or pixel group routed
through the switching matrix.
Typical \ac{IV} scans extend up to approximately 300~V, depending on
the sensor under test.

To protect the sensor during the bias sweep, the current compliance of
the \ac{smu} is used as a hardware safety limit.
When the measured current reaches the compliance threshold, the control
software immediately initiates a return sweep to 0~V.
This prevents excessive current flow near breakdown and shortens the
measurement time for channels that show an early current rise.

For \ac{CV} measurements, a Wayne Kerr
43100\footnote{\url{https://www.waynekerrtest.com/products_detail.php?indexs=7&brand=Wayne\%20Kerr}}
LCR meter is used, providing capacitance resolution down to
$10^{-5}$~pF and an internal DC bias of up to 40~V.
For measurements requiring a higher bias voltage, an external DC source
can be added in series between the switching matrix and the LCR meter.

\section{Software Design}\label{sec:software}

The characterization of large pixelated LGAD sensors, such as the
$16 \times 16$ array considered in this work, requires coordinated
control of the switching matrix and the precision measurement
instruments, together with automated scan execution and stable remote
operation.
To meet these requirements, dedicated control software was developed
with the following technical objectives:

\begin{itemize}
  \item Ensure synchronized control of the switching matrix and the
  measurement instruments, so that the electrical path
  configuration remains consistent with the measurement sequence.
  \item Support fully automated scan sequences over all 256 channels to
  minimize manual intervention and improve measurement consistency.
  \item Provide both a graphical user interface (GUI) for interactive
  monitoring and a command-line interface (CLI) for terminal-based
  operation and scripted automation.
  \item Enable stable remote access to the measurement system, allowing
  operation without direct physical proximity to the hardware setup.
\end{itemize}

\begin{figure}[h]
\begin{center}
\includegraphics[width=0.99\linewidth]{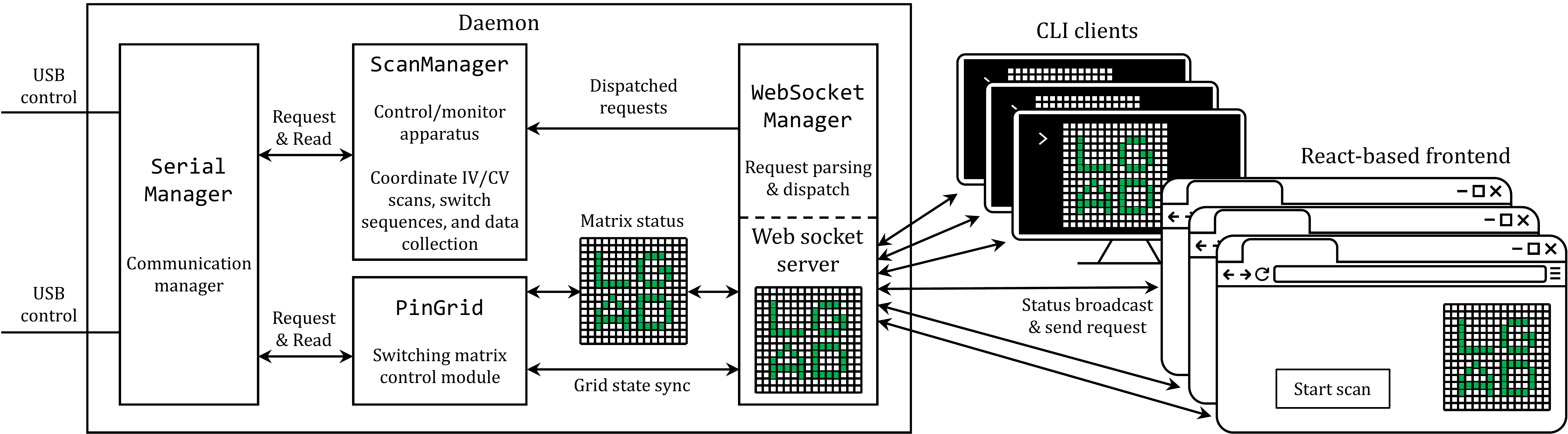}
\end{center}
\caption{Block diagram of the software architecture.
The server runs as a Linux daemon and consists of four main components:
\texttt{SerialManager}, which communicates with the switching hardware
and measurement instruments; \texttt{ScanManager}, which coordinates
automated \ac{IV} and \ac{CV} scans; \texttt{PinGrid}, which maintains
the switching-matrix state; and \texttt{WebSocketManager}, which handles
communication with connected clients.
Both GUI and CLI clients can monitor the system state and interact with
the server in real time.}
\label{fig:sw}
\end{figure}

To satisfy these requirements, the software was implemented using a
server-client architecture.
This decoupled design improves modularity and allows the user interface
to run on a different network node from the measurement server.
Communication between the server and the clients is implemented through
the WebSocket protocol using JSON-formatted messages.
The overall architecture of the software system, including the
interaction between the server and multiple clients, is illustrated in
Fig.~\ref{fig:sw}.

The server is implemented in C++ and runs as a background Linux daemon
to provide continuous access to the measurement system.
At startup, it automatically establishes communication with the
Raspberry Pi Pico and the connected measurement instruments.
Its core functionality is organized into four main software components:

\begin{itemize}
  \item \texttt{PinGrid}: maintains the logical state of the
  $16 \times 16$ switching matrix and keeps the software-defined channel
  configuration consistent with the physical switch states.
  \item \texttt{SerialManager}: handles low-level serial communication
  with the switching hardware and the external measurement instruments.
  \item \texttt{WebSocketManager}: manages client connections, parses
  incoming requests, and distributes updated system status information.
  \item \texttt{ScanManager}: coordinates automated \ac{IV} and \ac{CV}
  scan sequences, including predefined scan modes ranging from full-array
  scans to selected regions of interest.
\end{itemize}

The server supports simultaneous connections from multiple clients.
When more than one client is connected, the current system state,
including the switching-matrix configuration, is synchronized across all
active interfaces in real time.

Two complementary client interfaces were developed for interaction with
the server.
The GUI client is a web-based application built with the React
library\footnote{\url{https://react.dev}}, providing a visual
representation of the $16 \times 16$ pixel grid.
It allows users to monitor the real-time channel status, inspect the
switching configuration, initiate scan sequences, and view system logs.
The CLI client supports operation in environments without a graphical
desktop and is particularly useful for scripting and integration into
broader automated testing workflows.

\section{Performance}\label{sec:performance}
The developed system was validated through automated \ac{IV} and
\ac{CV} measurements of a $16 \times 16$ LGAD array.
In a typical measurement sequence, the probe card is first aligned to
the sensor and electrical contact is established.
A target pixel or pixel group is then selected through the switching
matrix, after which an \ac{IV} or \ac{CV} curve is acquired by a voltage
sweep.
The measured data are stored in plain text format, and the system then
proceeds to the next channel configuration by returning to the
channel-selection step.
This sequence is repeated until all designated channels are scanned.

\begin{figure}[htbp]
    \centering
    \begin{subfigure}[b]{0.56\textwidth}
        \centering
        \includegraphics[width=\textwidth]{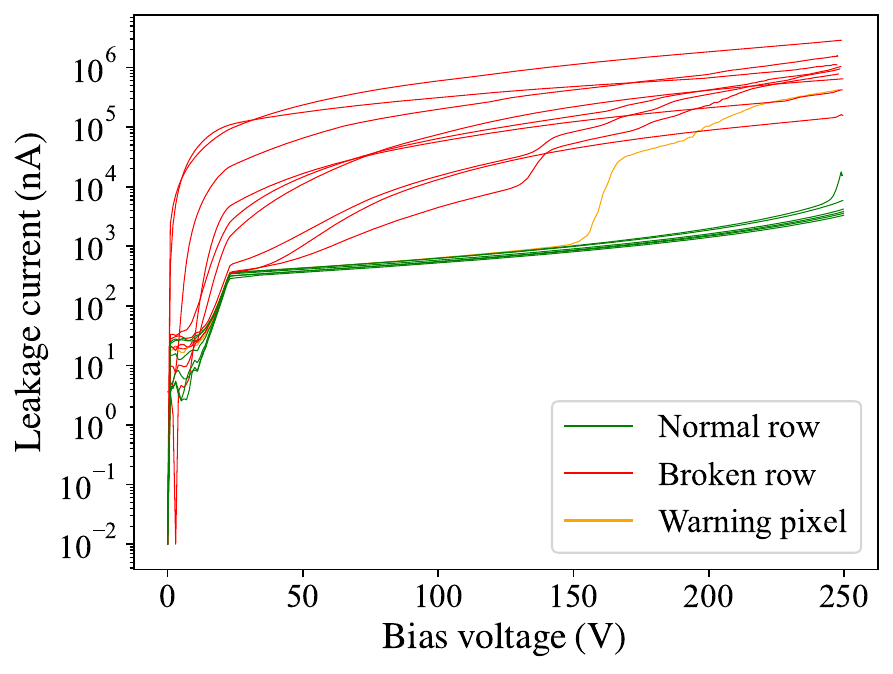}
        \caption{}
        \label{fig:iv_row_curves}
    \end{subfigure}
    \hfill
    \begin{subfigure}[b]{0.42\textwidth}
        \centering
        \includegraphics[width=\textwidth]{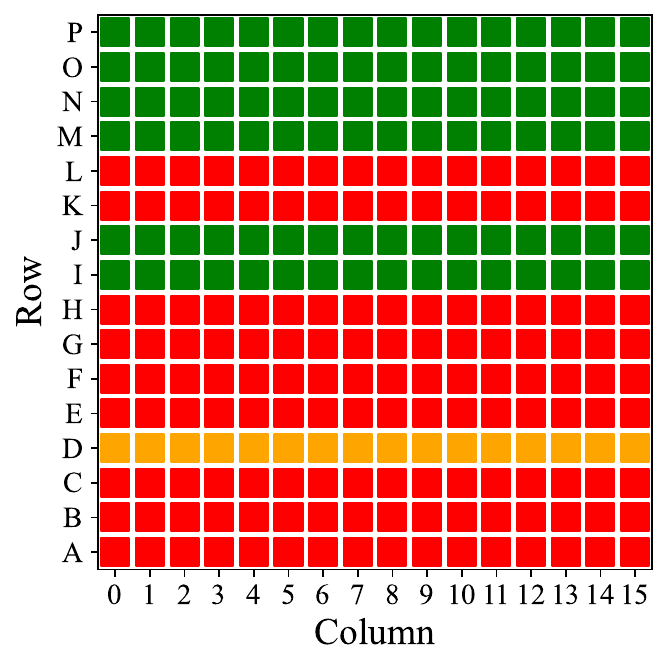}
        \caption{}
        \label{fig:iv_row_map}
    \end{subfigure}
    \caption{\ac{IV} measurement results from the row-wise scan of a
    $16 \times 16$ LGAD array.
    (a) \ac{IV} curves measured for each row, with rows classified as
    normal (green), warning (orange), or broken (red) based on the
    breakdown behavior of the pixels within the row.
    (b) Row-status map of the full array.
    Rows are classified as broken if they contain one or more pixels with
    significant premature breakdown, and as warning if they contain pixels
    with moderately early breakdown.}
    \label{fig:iv_row}
\end{figure}

As a rapid pre-screening step, row-wise \ac{IV} measurements were first
performed by connecting all pixels within a row simultaneously.
With this configuration, the full $16 \times 16$ array could be scanned
in approximately 20 minutes, providing a quick overview of the
row-level \ac{IV} of the sensor.
Figure~\ref{fig:iv_row}(a) shows the resulting \ac{IV} curves for each
row, and Fig.~\ref{fig:iv_row}(b) shows the corresponding row-status
map.
Rows composed entirely of functional pixels are shown in green, rows
containing pixels with moderately early breakdown are shown in orange,
and rows containing one or more pixels with significant premature
breakdown are shown in red.

\begin{figure}[htbp]
    \centering
    \begin{subfigure}[b]{0.56\textwidth}
        \centering
        \includegraphics[width=\textwidth]{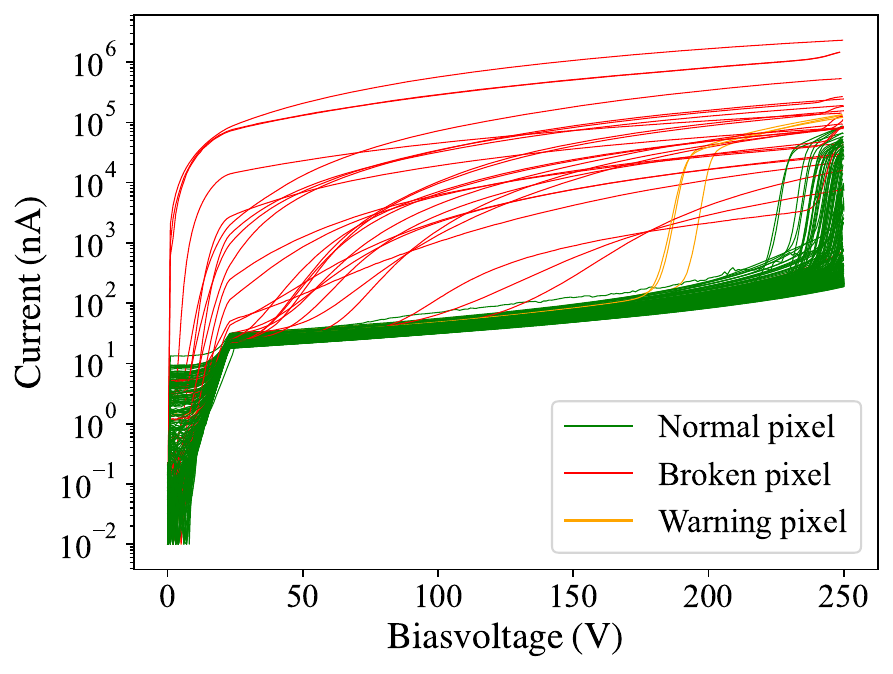}
        \caption{}
        \label{fig:iv_pad_curves}
    \end{subfigure}
    \hfill
    \begin{subfigure}[b]{0.42\textwidth}
        \centering
        \includegraphics[width=\textwidth]{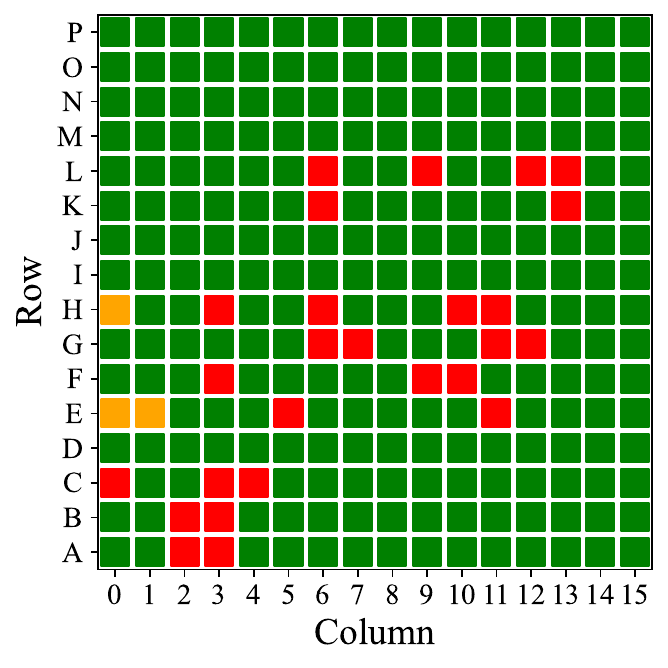}
        \caption{}
        \label{fig:iv_pad_map}
    \end{subfigure}
    \caption{\ac{IV} measurement results from the pixel-by-pixel scan of a
    $16 \times 16$ LGAD array.
    (a) \ac{IV} curves measured for all pixels, with pixels classified as
    normal (green), warning (orange), or broken (red) according to their
    breakdown behavior relative to the designed operating voltage of
    approximately 250~V.
    (b) Pixel-status map of the full array showing the spatial distribution
    of each pixel category.}
    \label{fig:iv_pad}
\end{figure}

Following the row-wise pre-screening, pixel-by-pixel \ac{IV}
measurements were performed for detailed characterization of the full
array.
For a voltage sweep from 0 to 300~V with a 1~V step, the full
$16 \times 16$ scan required approximately 340 minutes.
Figure~\ref{fig:iv_pad}(a) shows the \ac{IV} curves measured for all
pixels, and Fig.~\ref{fig:iv_pad}(b) shows the corresponding
pixel-status map.
Each pixel is classified into three categories based on its breakdown
behavior relative to the designed operating voltage of approximately
250~V: functional pixels are shown in green, pixels with moderately
early breakdown are shown in orange, and pixels with significantly
premature breakdown are shown in red.

This two-stage approach demonstrates a practical measurement strategy
for large-scale LGAD quality control.
The rapid row-wise scan, completed in approximately 20 minutes,
provides an efficient pre-screening step for identifying rows that
require closer inspection.
A subsequent pixel-by-pixel scan can then be focused on the flagged
rows, rather than being applied to the entire array from the outset.
For the measurement conditions used here, this significantly improves
the efficiency of sensor characterization compared with performing a
full-array pixel-by-pixel scan alone.

\begin{figure}[htbp]
    \centering
    \includegraphics[width=0.7\linewidth]{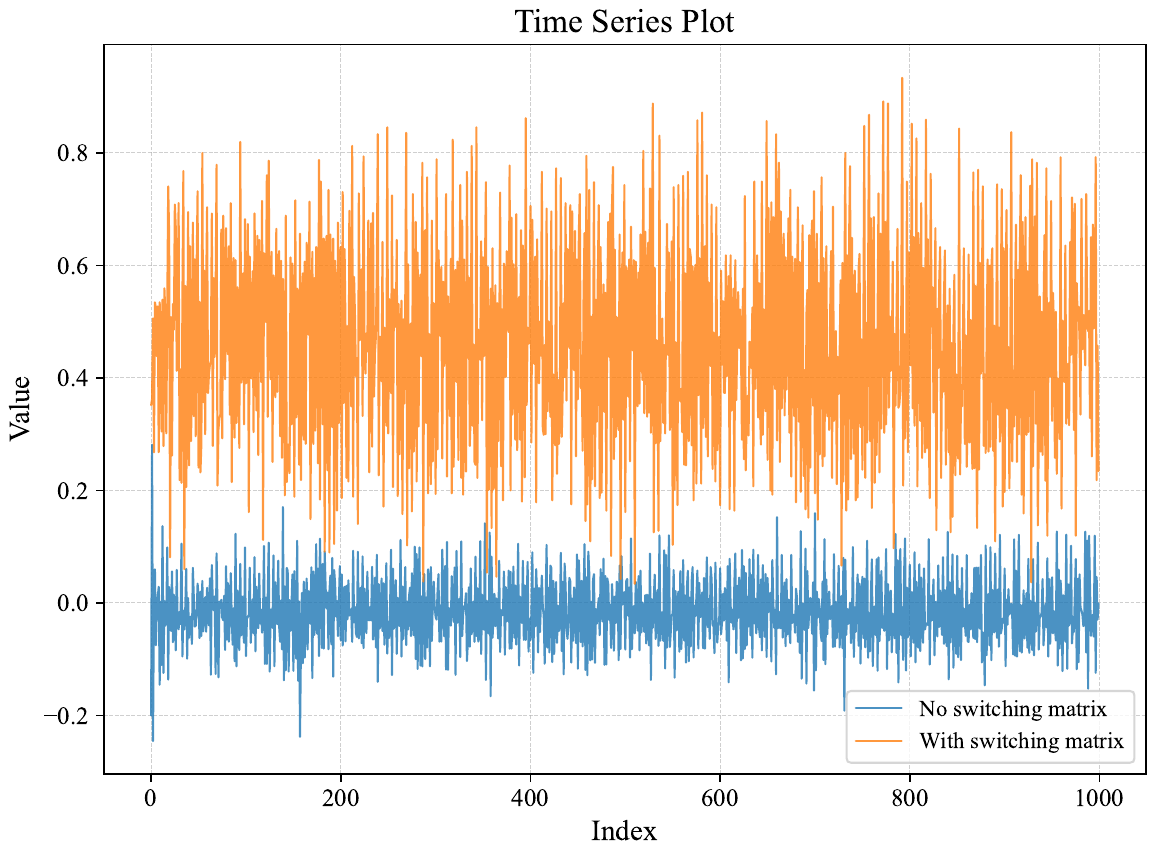}
    \caption{Comparison of the measured leakage current with and without the
    switching matrix in the readout path.
    The picoammeter was connected directly to a single LGAD pixel in the
    reference configuration (blue), and the same signal was routed through
    the switching matrix in the switched configuration (orange).
    The switching matrix introduces an offset of approximately 0.46~nA and
    increases the standard deviation from 0.066~nA to 0.19~nA.}
    \label{fig:noise_comparison}
\end{figure}

To evaluate the offset and noise introduced by the switching matrix, the
leakage current of a single LGAD pixel was measured both with and
without the switching matrix in the readout path.
In the reference configuration, the picoammeter was connected directly
to the pixel.
In the switch configuration, the same signal was routed through the
switching matrix.
Figure~\ref{fig:noise_comparison} shows the measured current as a
function of time for both cases.
Without the switching matrix, the measured current has a standard
deviation of 0.066~nA.
With the switching matrix, an offset of approximately 0.46~nA is
introduced and the standard deviation increases to 0.19~nA.
Even in the extreme case where all 256 outputs are connected to a
single readout channel, the total leakage current contribution from the
switching matrix remains below 1~nA.
These values remain small compared with the typical leakage current of a
normal LGAD pixel, indicating that the switching matrix does not
significantly degrade the measurement quality.

\begin{figure}[htbp]
    \centering
    \begin{subfigure}[b]{0.56\textwidth}
        \centering
        \includegraphics[width=\textwidth]{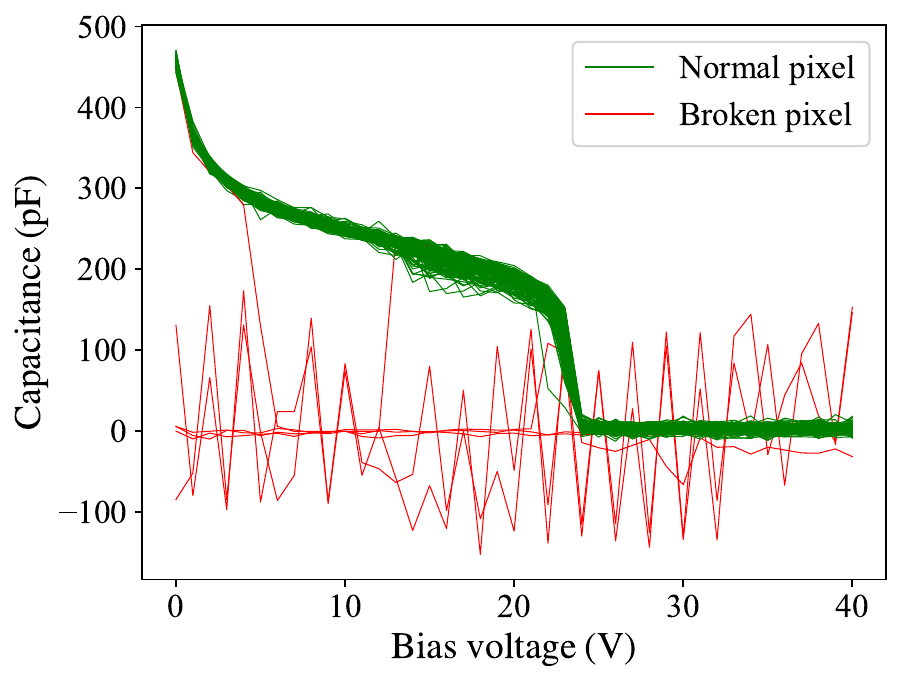}
        \caption{}
        \label{fig:cv_curves}
    \end{subfigure}
    \hfill
    \begin{subfigure}[b]{0.42\textwidth}
        \centering
        \includegraphics[width=\textwidth]{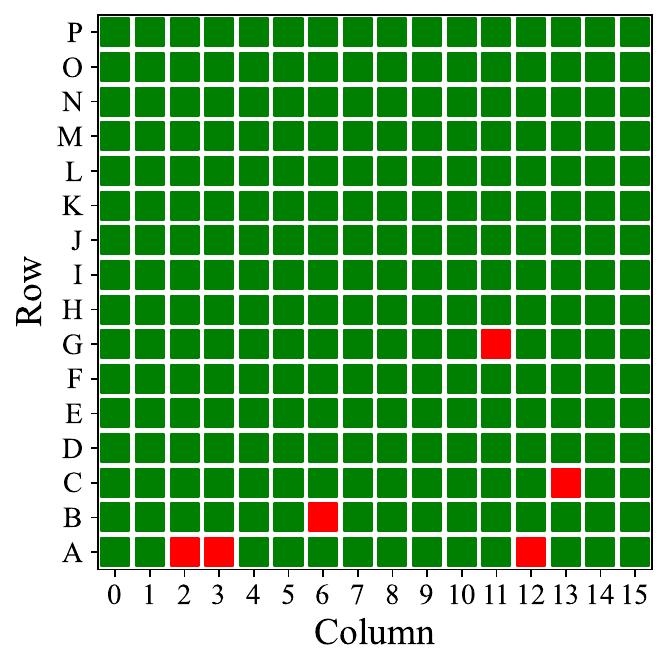}
        \caption{}
        \label{fig:cv_map}
    \end{subfigure}
    \caption{\ac{CV} measurement results from the pixel-by-pixel scan of a
    $16 \times 16$ LGAD array.
    (a) \ac{CV} curves measured for all pixels, with pixels classified as
    normal (green) or broken (red) according to the presence or absence of
    a well-defined depletion transition.
    (b) Pixel-status map of the full array showing the spatial distribution
    of the two categories.}
    \label{fig:cv_ex}
\end{figure}

Pixel-by-pixel \ac{CV} measurements were also performed for the same
sensor.
For a bias sweep up to 40~V, each channel required approximately one
minute, with the scan speed primarily limited by the performance of the
LCR meter.
Figure~\ref{fig:cv_ex}(a) shows the \ac{CV} curves measured for all
pixels, and Fig.~\ref{fig:cv_ex}(b) shows the corresponding pixel-status
map.
Normal pixels exhibit consistent \ac{CV} characteristics with a clear
depletion transition, whereas broken pixels do not show a well-defined
\ac{CV} curve and instead exhibit large capacitance fluctuations over
the full bias range.

The pixel-status maps obtained from the \ac{IV} and \ac{CV}
measurements do not necessarily coincide, because the two measurements
probe different electrical characteristics of the sensor.
In particular, the \ac{CV} scan in this study was limited to 40~V,
whereas the \ac{IV} scan was used to evaluate breakdown behavior near
the operating voltage.
As a result, pixels showing premature breakdown in the \ac{IV}
measurement can still exhibit a well-defined \ac{CV} response within the
lower voltage range covered by the \ac{CV} scan.

\section{Conclusions}\label{sec:summary}

We have developed and validated an automated probe card system for the
electrical characterization of full-size pixelated LGAD sensors.
The system is designed for $16 \times 16$ LGAD arrays and supports
automated row-wise and pixel-by-pixel \ac{IV} and \ac{CV}
measurements, addressing the need for reliable and scalable quality
control in high-granularity timing-detector applications.

The developed platform combines a pogo-pin probe card matched to the
sensor geometry, a 256-channel modular switching matrix, precision
measurement instruments, and integrated control software based on a
server-client architecture.
The switching system enables selective access to individual pixels while
holding non-selected pixels at ground, and also supports multi-pixel
row-wise measurement configurations.
The additional offset and leakage introduced by the switching matrix
remain small compared with the leakage current of a normal LGAD pixel,
showing that the switching architecture does not significantly degrade
the measurement quality.

Automated measurements of a $16 \times 16$ LGAD array were successfully
demonstrated with the system.
A row-wise \ac{IV} scan of the full array was completed in
approximately 20 minutes, while a pixel-by-pixel \ac{IV} scan from
0 to 300~V with a 1~V step required about 340 minutes.
This enables a practical two-stage workflow in which rapid
pre-screening is followed by a detailed inspection of flagged regions,
thereby improving the efficiency of large-scale sensor
characterization.

These results demonstrate that the developed system provides a practical
and scalable solution for LGAD quality control.
Its modular hardware design, flexible switching architecture, and
automation capability make it well-suited for use in distributed sensor
testing environments and future large-scale LGAD production and
evaluation campaigns.

\acknowledgments

This work is supported by the National Research Foundation of Korea (NRF) grants funded by the Korea government (MSIT) (RS-2008-NR007227, RS-2020-NR050645, and RS-2025-00560964) and a Korea University Grant.

\bibliographystyle{unsrtnat}
\bibliography{refs.bib} 

\end{document}